\documentclass{ws-procs975x65}
\usepackage{epstopdf}

\begin{document}

\title{Magnetic Reconnection Instabilities in Soft-Gamma Repeaters}
\author{Jeremy S. Heyl and Ramandeep Gill}
\address{Department of Physics and Astronomy, University of British
  Columbia, Vancouver BC Canada}
\begin{abstract}
  We examine an external trigger mechanism that gives rise to the intense
  soft gamma-ray repeater (SGR) giant flares. Out of the three giant
  flares, two showcased the existence of a
  precursor, which we show to have had initiated the main flare. We
  develop a reconnection model based on the hypothesis that shearing
  motion of the footpoints causes the materialization of a
  Sweet-Parker current layer in the magnetosphere. The thinning of
  this macroscopic layer due to the development of an embedded
  super-hot turbulent current layer switches on the impulsive Hall
  reconnection, which powers the giant flare. We show that the
  thinning time is on the order of the pre-flare quiescent time.
\end{abstract}

\section{Preliminaries}

We take the observation that two of the three giant flares from the
SGRs were preceded by a precursor that was similar in energy ($\sim
10^{41}~\text{erg}$) to a typical short SGR burst (Ibrahim et
al. 2001; Hurley et al. 2005) as a hint that the precursor ``lights
the fuse'' for the giant flare.  The natural timescale for this fuse
is the Alfv\'en time of the inner magnetosphere, which for
exceptionally low values of the plasma beta parameter is very small;
$\tau_A\sim R_{\star}/c\sim 30 \mu$s.  Although this is close to the
rise time of the flare, this timescale is up to six orders of
magnitude shorter than the delay between the precursor and the flare.
Of course, the symmetries of the magnetic field coupled to the plasma
prevent the quasi-steady-state configuration to change this quickly
unless the gradient of the magnetic field are large. So, the questions
are how does the field develop large gradients and can this happen on
the timescale of the delay between the precursors and the giant
flares.

\begin{figure}
\epsfig{file=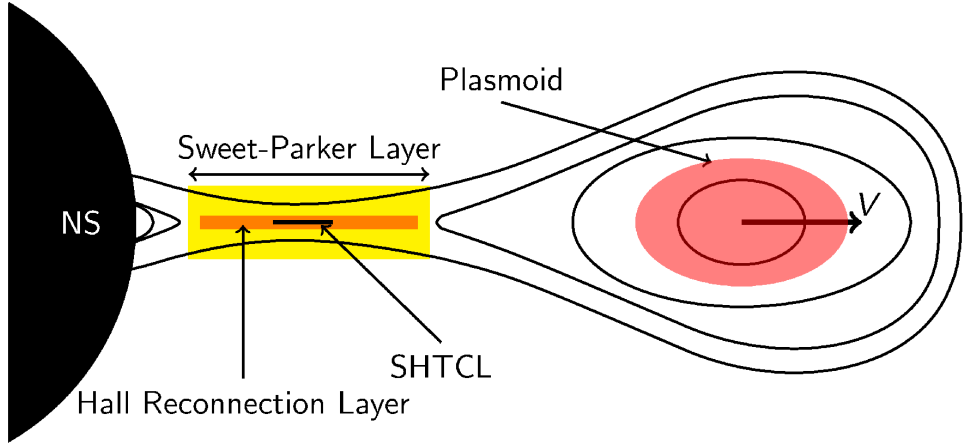,width=\linewidth,clip=} 
\caption{This figure displays the setup of the different reconnecting
  current layers.  The macroscopic Sweet-Parker layer with length
  $L\sim10^5$ cm and width $\delta\sim0.01$ cm is the largest of the
  three. This layer is then thinned down vertically as strong magnetic
  flux is convected into the dissipation region. The Hall reconnection
  layer, represented by the dark gray region, develops when $\delta$
  becomes comparable to the ion-inertial length $d_i$. The system
  makes a transition from the slow to the impulsive reconnection and
  powers the main flare. The tiny region embedded inside the
  Sweet-Parker layer is the super-hot turbulent current layer, which
  aids in creating sufficient anomalous resistivity to facilitate the
  formation of the Sweet-Parker layer. The strongly accelerated plasma
  downstream of the reconnection layer is trapped inside magnetic flux
  lines and forms a plasmoid moving at some speed $V$. This plasmoid
  is then finally ejected during the initial spike when the external
  field undergoes a sudden relaxation (After Lyutikov 2006).}
\label{fig:NSfig}
\end{figure}

Figure~\ref{fig:NSfig} outlines a scenario where the initial SGR flare
which is associated with a crustal shift injects a twist in the 
magnetosphere (Thompson \& Duncan 1995). This causes the external magnetic field lines to form a configuration 
allowing the formation of a current sheet where the field lines
can reconnect and release a large amount of magnetic energy 
(see for e.g. Lyutikov 2006).
\begin{figure}
\epsfig{file=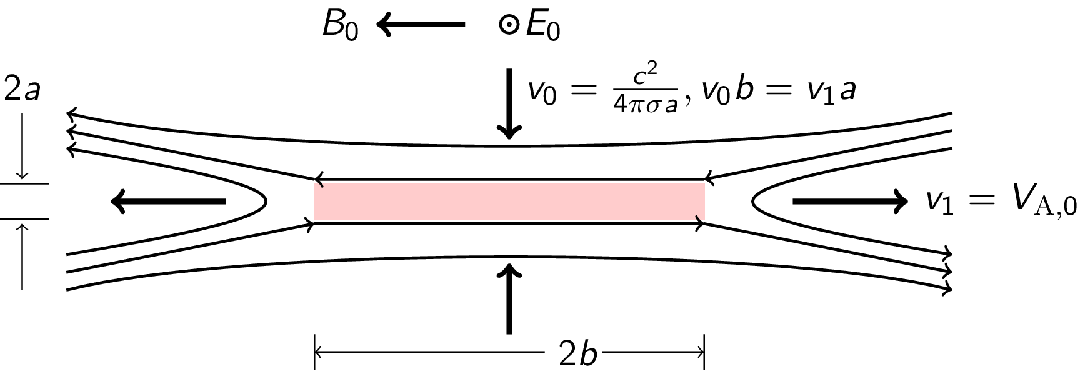,width=0.48\linewidth,clip=} \hfill
\epsfig{file=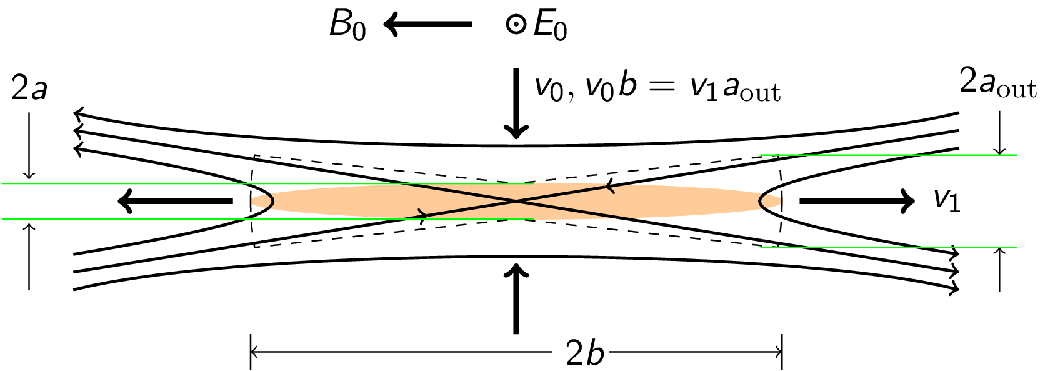,width=0.48\linewidth,clip=} 
\caption{This figure focusses on the two reconnection regions.  The
  left panel is the Sweet-Parker layer, and the right panel is the
  Super-Hot Turbulent-Current Layer (SHTCL).  See Somov (2006) for
  further details.
}
\label{fig:current_layers}
\end{figure}

This reconnection will proceed slowly over a resistive timescale.  The
standard picture for this process is the Sweet-Parker layer denoted in
Figure~\ref{fig:NSfig} and placed into focus in the left panel of
Figure~\ref{fig:current_layers}.  We assume that the plasma is
incompressible and that the magnetic energy is converted to kinetic
energy. Combining these assumptions with the geometry then yields 
a relationship between the Alfv\'enic Mach number of the reconnecting 
flow and the resistivity through the Lundquist number,
\begin{equation}
M_\mathrm{A} =
\frac{v_0}{V_{\mathrm{A},0}} \sim \left ( \frac {c^2}{4\pi\sigma b}
\right )^{1/2} = N_L^{-1/2}.
\end{equation}
Because the Spitzer resistivity of the collisionless plasma at
temperatures $\sim 10^8~\text{K}$ is small, this Mach number is small,
and the reconnection timescale is large unless some sort of anomalous
resistivity is present.  We will use the super-hot turbulent current
layer (SHTCL) model to introduce a source of anomalous resistivity
(Somov 2006).  Again making the same assumptions as with the
Sweet-Parker layer but with the modified geometry of the right panel
of Figure~\ref{fig:current_layers}, we obtain the reconnection
velocity of
\begin{equation}
M_\mathrm{A} =
\frac{v_0}{V_{\mathrm{A},0}} \sim \frac{a_\mathrm{out}}{b} \sim
\frac{B_\perp}{B_0}
\end{equation}
which now depends on geometric considerations rather than the Spitzer
resistivity of the plasma.

What length scales do we have?  From biggest to smallest, we have the
layer width ($b$), the layer funnel width ($a_\mathrm{out}$), the
layer thickness ($a$), the proton cyclotron radius, and the electron
cyclotron radius. Cassak et al. (2006) show that the Sweet-Parker layer 
can be thinned down to proton cyclotron lengthscale by convecting 
strong fields into the current layer. We find the critical field strength 
for this to occur $B_c\sim10^{14}~\text{G}$. With the presence of baryons 
the thinned current layer undergoes Hall reconnection which proceeds on an Alfv\'enic timescale,
hundreds of microseconds typically. Thus, the delay between the precursor
to the flare is the timescale to thin from the initial reconnection
region (this determines the total energy of the flare) to the proton
cyclotron radius.
\section{Conclusions}
The thin SHTCL provides anomalous resistivity that allows the current
layer to thin on a timescale comparable to the delay for the 
December 27 event
\begin{eqnarray}
\tau_\mathrm{thin} & \sim & 2W_s\sqrt{\frac{L}{\eta_{\rm diff}c}\left(\frac{B_c}{B_0}
\right)} \\
& \sim & 130\mbox{ s }\left(\frac{W_s}{10^5\mbox{ cm}}\right)\left(
\frac{L}{10^5\mbox{ cm}}\right)\left(\frac{B_0}{10^{14}\mbox{ G}}
\right)^{-1/2} \!\!\! \left(\frac{n_b}{6\times10^{22}\mbox{ cm}^{-3}}\right)^{3/4}
\end{eqnarray}
where $W_s$ is the magnetic field shear length, $L$ is the length of the current sheet, 
\begin{equation}
n_b \sim 6\times10^{22}\left(\frac{E_\mathrm{th}}{10^{38}\mathrm{ erg}}\right)
\left(\frac{R_{\star}}{10^6\mathrm{ cm}}\right)^{-2}\left(\frac{M_{\star}}
{1.4\mathrm{M}_{\odot}}\right)\mathrm{ cm}^{-3}
\end{equation}
We have used the data for the December 27 event.  The August 27 flare
had a short and weak precursor, yielding about one percent of the
baryons. Furthermore, the flare was weaker too, requiring a smaller
reconnection region. We get
\begin{eqnarray}
\tau_\mathrm{thin} & \sim & 0.4\mbox{ s }\left(\frac{W_s}{2 \times
    10^4\mbox{ cm}}\right)\left(\frac{L}{5 \times 10^4\mbox{ cm}}\right)\left(\frac{B_0}{10^{14}\mbox{ G}}
\right)^{-1/2} \!\!\!
\left(\frac{n_b}{6\times10^{20}\mbox{ cm}^{-3}}\right)^{3/4}
\end{eqnarray}
similar to the delay for the August 27 event.  Future giant flares from
SGRs will improve our understanding of such correlations as well as 
the mechanism for the triggering of giant flares from soft-gamma repeaters.
\section*{Bibliography}
Cassak P. A., Drake J. F., Shay M. A., 2006, ApJL, 644, L145\\
Gill R., Heyl J. S., 2010, MNRAS, 407, 1926 \\
Hurley K., et al., 2005, Nature, 434, 1098 \\
Ibrahim A. I., et al., 2001, ApJ, 558, 237 \\
Lyutikov M., 2006, MNRAS, 367, 1594\\ 
Somov, B., 2006, {\em Plasma Astrophysics, Part II : Reconnection and Flares}, Astrophysics and space science library (ASSL), Vol. 341. Dordrecht: Springer\\
Thompson C., Duncan R. C., 1995, MNRAS, 275, 255

\end{document}